# Aquifer: Hierarchical Memory Pooling with CXL and RDMA for MicroVM Snapshots

Junliang Hu Huaicheng Li[#] Ming-Chang Yang

The Chinese University of Hong Kong
Department of Computer Science and Engineering

[#]Virginia Tech
Department of Computer Science

## Abstract

Memory stranding wastes 25–35% of installed DRAM in production cloud clusters. Memory pooling over CXL and RDMA offers a remedy, but neither technology alone suffices: CXL provides low-latency, load/store-transparent access limited to a pod, while RDMA provides cluster-wide reach at higher latency with software overhead. A hierarchical architecture combining both tiers is the practical path forward, yet remains unexplored for MicroVM-based serverless computing, where snapshot restore latency is the dominant cold-start bottleneck.

We present Aquifer, the first system to serve MicroVM snapshots from a hierarchical CXL+RDMA memory pool. A characterization of snapshot images reveals that the vast majority of pages are either zero or cold, enabling a hotness-based snapshot format that eliminates zero pages and places only the hot working set in the CXL pool while storing cold pages in the RDMA pool. Sharing these snapshots across hosts on CXL 2.0 multi-headed devices, which lack hardware cache coherence, requires Aquifer's ownership-based coherence protocol to ensure correctness. Finally, Aquifer uses a copy-based page serving mechanism pre-installs hot pages from CXL memory before MicroVM resume and demand-pages cold pages asynchronously from RDMA. On emulated CXL+RDMA hardware, Aquifer achieves a 2.2 × geometric-mean speedup in end-to-end invocation time over Firecracker and 1.1× over the next best alternative.

## 1 Introduction

Memory stranding, where provisioned DRAM sits idle on a server because the co-located CPU cores are already fully subscribed, is a persistent and costly problem in modern cloud data centers. [21, 22, 31] Measurements at Microsoft Azure show that stranding can waste 25–35% of installed memory capacity across production clusters, and the problem worsens as utilization increases [35]. Similar observations have been reported at Google [31] and Meta [47], and recent characterizations at Alibaba confirm that 45% of memory remains stranded across their fleet on average [17]. At warehouse scale, even single-digit percentage improvements in memory utilization translate to hundreds of millions of dollars in DRAM savings.

Memory pooling, decoupling memory from individual servers and making it accessible over a shared fabric, is the natural remedy. Early work explored pooling over RDMA, [8, 11, 23, 42] but RDMA-based disaggregation incurs software overhead on every remote access: page faults, kernel intervention, or application-level smart pointers [42] are required to trap and redirect memory references.

The Compute Express Link (CXL) standard [1] promises a fundamentally different paradigm in this design space. By extending the processor's load/store interface over a PCIe-based interconnect, CXL enables remote memory to appear as a memory only NUMA node that the CPU can load/store transparently, without any software interposition on the data path [1]. This hardware-level integration makes CXL a compelling substrate for practical memory pooling in the cloud [35]. However, real CXL deployments [49] differ significantly from the idealized picture.

CXL multi-headed memory devices (MHDs) [13, 25, 28, 29, 35] can expose a single memory expander to a small number of hosts, but scalability is limited to a pod of 8–16 sockets [29], and full hardware cache coherence across hosts is either absent [49] or only partially supported [29]. Extending pooling beyond a single pod requires a second tier of connectivity, typically RDMA over a cluster-wide Clos network, with different access pattern and performance characteristics. The performance cost [36] and signal integrity [35] of CXL switches further limit the scope of any single coherence domain. These realities suggest that a hierarchical pooling architecture, which combines a fast, pod-local CXL tier with a larger, cluster-wide RDMA tier, is the most practical path toward flexible memory disaggregation. Yet this architecture remains largely unexplored.

Meanwhile, MicroVM-based [14] serverless computing has become a major service model offered by cloud providers. Platforms such as AWS Lambda, backed by the Firecracker MicroVM monitor [7], run millions of function invocations per second. A central challenge in these systems is cold-start latency: when a function has no warm instance available, the platform must restore a full MicroVM from a serialized snapshot, loading guest kernel state, application memory, and device models, before the function can begin executing. This process is both latency-critical (users expect sub-second startup) and storage-intensive (snapshots for each function version must be persisted and made accessible to any server in the cluster). Snapshot-based warm restore has been the focus of recent works [12, 46].

Prior systems that accelerate MicroVM restore have focused exclusively on single-machine optimizations: demand-paging snapshots from local or network-attached storage, and prefetching based on recorded access traces [12, 46]. These approaches treat each server as an island and worsen resource stranding. A separate line of work has begun to explore memory pooling for serverless containers: CXLFork [10] uses CXL pool as a shared memory primitive to clone process states across nodes. FaaSMem [48] swaps out idle memory to remote RDMA pool. TrEnv [28] offloads application specific memory to a remote pool consists of either RDMA or CXL memory. However, these systems assume a flat, homogeneous pool, do not address the heterogeneity of a hierarchical CXL+RDMA architecture, and target cgroup-based containers rather than MicroVMs sandboxes with their distinct snapshot restore semantics.

In this paper, we present Aquifer, the first system to exploit a hierarchical memory pool architecture for MicroVM-based serverless computing. Aquifer combines a pod-local CXL memory pool, realized through multi-headed CXL devices, with a cluster-wide RDMA memory pool interconnected over a conventional Clos network. Function snapshots are distributed across these tiers according to access frequency and latency sensitivity: hot pages reside in the CXL pool for sub-microsecond serving during restore, while the cold subsets is backed by the RDMA pool to harvest stranded memory. This hierarchical design allows Aquifer to achieve a 2.2× geometric-mean speedup in end-to-end invocation time over Firecracker and 1.1× over REAP, the next-best alternative, across nine serverless workloads.

Realizing this architecture requires addressing several key challenges that do not arise in single-machine or flat pooling designs. First, snapshot storage and data placement: full VM memory images are dominated by zero and cold pages that waste pool capacity; Aquifer must design a compact snapshot format that eliminates redundancy and partitions the remaining pages across the CXL and RDMA tiers according to access hotness. Second, cache coherence: existing CXL 2.0 MHDs provide no coherence guarantees [49] across hosts sharing a pool; Aquifer must ensure that concurrent snapshot accesses by different hosts see consistent state without relying on hardware cache coherence. Third, page installation: the compact, non-file-backed snapshot format precludes conventional `mmap`-based page serving, and the hot set is highly fragmented across the guest address space; Aquifer must efficiently install snapshot pages into the guest address space while keeping the shared pool image immutable across concurrent restores.

We make the following contributions:

- A characterization of MicroVM snapshot image composition across 9 representative serverless workloads, revealing that 82.8% of pages are zero and the hot working set constitutes only 5.5% of total pages, with highly fragmented address layout. These findings motivate a compact, hotness-aware snapshot format and a copy-based page installation mechanism (§2.3.3 - §2.3.4).
- The design and implementation of Aquifer, a hierarchical CXL+RDMA memory pooling system for MicroVM snapshot serving that integrates with the Firecracker MicroVM monitor. Aquifer introduces (1) a hotness-based snapshot format that eliminates zero pages and places hot pages in the CXL pool and cold pages in the RDMA pool; (2) an ownership-based coherence protocol that ensures correct concurrent snapshot access on CXL 2.0 MHDs without hardware cache coherence; and (3) a copy-based page serving mechanism that pre-installs hot pages from CXL memory before MicroVM resume and demand-pages cold pages asynchronously from the RDMA pool (§3 - §4).
- An evaluation on CloudLab [20] hardware showing that Aquifer achieves a 2.2× geometric-mean speedup in end-to-end invocation time over Firecracker and 1.3× over FaaSnap across nine serverless workloads, while scaling to 32 concurrent MicroVM restores (§5).

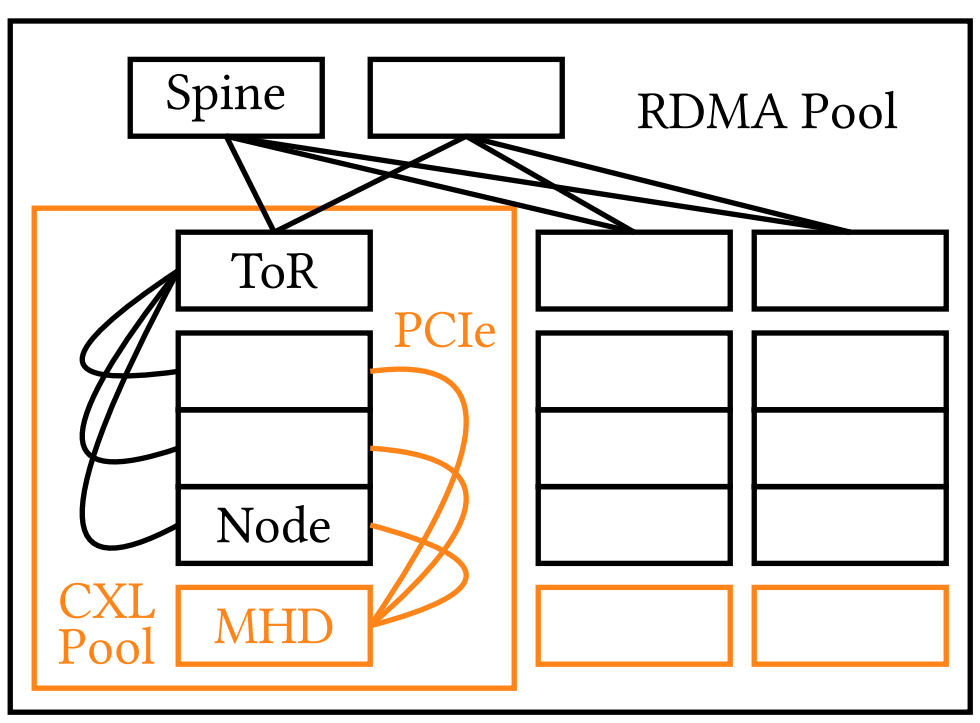

Figure 1: Hierarchical memory pool architecture with pod-level CXL and cluster-level RDMA memory pools.

## 2 Background and Motivation

### 2.1 Memory Disaggregation and Pooling

**RDMA-based memory disaggregation.** RDMA over Converged Ethernet (RoCE) is already deployed in virtually all major cloud data centers [24], making the entire data center network RDMA-capable and, in principle, able to form a unified cluster-wide memory pool. Remote memory regions exposed via RDMA can be accessed through two-sided operations (send/receive), which require participation from both endpoints, or one-sided operations (read/write), which allow an initiator to directly access remote memory without involving the remote CPU. One-sided RDMA operations are particularly attractive for memory pooling because they enable transparent harvesting of stranded memory on remote nodes without the remote node's participation. However, RDMA-based disaggregation still incurs software overhead on every remote access. Kernel-based remote swapping systems such as Infiniswap [23] and Fastswap [11] swap out pages to a remote memory pool, but when a swapped-out page is needed, a page fault must be handled and an RDMA read issued to fetch the page back, adding latency on the critical path. Application-integrated designs like AIFM [42] take a different approach: they use smart pointers to make individual objects offload-able to remote memory, but when an offloaded object is accessed, the runtime must intercept the reference and fetch the data from the remote node. In both cases, every remote access requires explicit software intervention to redirect the memory reference.

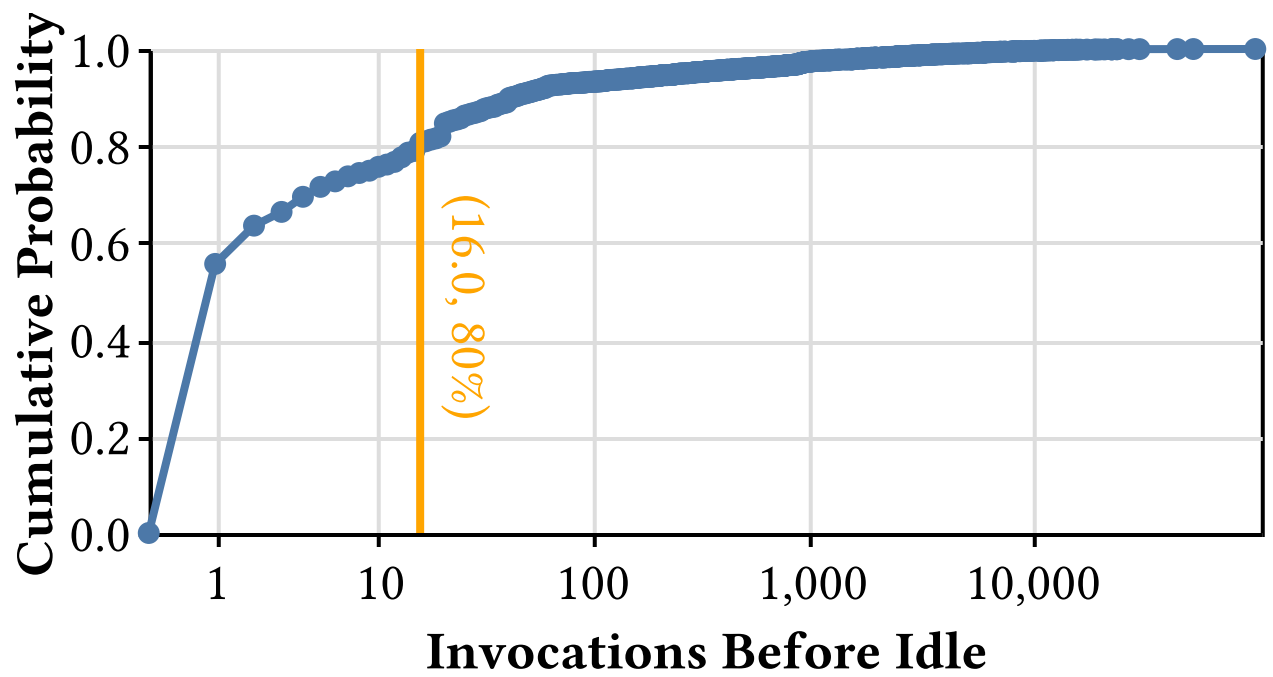

Figure 2: Distribution of consecutive invocation streak lengths before a function goes idle across all functions in the Azure Functions trace [50], measured over a two-week period with a 10-minute idle threshold.

**CXL-based memory pooling.** The Compute Express Link (CXL) standard [1] is an emerging contender that removes this software redirection entirely. CXL extends the processor's load/store interface over a PCIe-based interconnect, enabling remote memory to appear as a memory-only NUMA node that the CPU can access transparently [1]. Memory pooling support was introduced in CXL version 2.0, which is now supported by the newest CPU models, including the Intel Xeon 6 series and AMD's Turin-based lineup. CXL standard also specifies multi-headed devices (MHDs) [1], which allow a single CXL memory device to be shared among multiple server hosts. SK Hynix has demonstrated its Niagara 2.0 MHD [2], which runs CXL 2.0 and allows up to 8 hosts to share 4 channels of memory totaling up to 1 TiB. Although the CXL 2.0 specification formally supports only static, disjoint partitioning of physical memory across hosts, real devices found in industry [49] support sharing the same memory region among multiple hosts. This opens a new paradigm for memory pooling: connecting multiple hosts to a single MHD enables direct load/store access to shared pool memory at latencies of hundreds of nanoseconds, comparable to accessing a remote NUMA node [35, 36].

### 2.2 Serverless System

Serverless computing, also known as Function-as-a-Service (FaaS), is a modern cloud compute abstraction that provides finer resource granularity than traditional virtual machine or container hosting [9]. Tenants submit individual functions, and the platform dynamically allocates isolated sandboxes to execute each invocation on demand. This model increases elasticity for tenants, who pay only for actual execution time, and enables providers to flexibly schedule invocations across servers to improve overall resource utilization. However, when no pre-initialized sandbox is available for an incoming invocation, the platform must create one from scratch, a process known as *cold start*. Cold starts involve creating the sandbox, loading the runtime and application libraries, and initializing function state, which can take seconds and dominate the sub-second execution time of typical functions [12, 46].

**Cold start mitigation through keep-alive.** The most straightforward mitigation is to keep sandbox instances alive after an invocation completes, so that subsequent invocations can reuse a warm instance and skip the boot process entirely. Cloud providers commonly set the keep-alive period of a function to 10 minutes [43, 48]. Figure 2 shows the distribution of consecutive invocation streak lengths before a function goes idle in the Azure Functions trace [50]: 80% of instances receive 16 or fewer invocations per keep-alive window, meaning the vast majority of kept-alive instances serve only a handful of requests before becoming idle.

This creates severe memory pressure: even though the CPU is idle, the full memory state of each kept-alive instance must be retained in DRAM, reducing the capacity available for new invocations and creating large numbers of zombie sandboxes. FaaSMem [48] addresses this by partially offloading the memory of idle containers to a remote memory pool via RDMA [11, 23]. TrEnv [28] takes a different approach: it enables different functions to share the same kept-alive sandbox by categorizing memory state into *reusable* state (e.g., runtime and shared libraries) and *application-specific* state. The application-specific state is offloaded to a remote memory pool backed by either CXL or RDMA memory. When a new function invocation arrives, TrEnv loads the corresponding application-specific memory from the pool into the repurposed sandbox, avoiding a full cold start while amortizing sandbox creation cost across functions.

**Forking an alive cgroup-based sandbox.** Serverless sandboxes are either cgroup-based (i.e., containers) or MicroVM-based, and the cold-start mitigation strategy differs accordingly. For cgroup-based sandboxes, which are normal user-space processes, *fork*-based cloning [9] is a natural solution: forking creates a new process that inherits the initialized state of the parent, including loaded libraries and runtime data structures, thereby saving the initialization latency. CXLFork [10] advances this approach by decoupling the parent process state from the originating OS instance. Instead of relying on the OS-provided implicit state sharing through copy-on-write page table mappings between par-ent and child processes, CXLFork copies the process state to shared CXL memory and rebases OS-managed structures (e.g., page tables) to reference CXL physical addresses. This makes the saved state available for concurrent use by other OS instances on different nodes through direct remapping, enabling cluster-wide process cloning.

However, forking still require long kept alive time in order for the new invocations to enjoy the benefit of forking. As the child states are at least partially cloned from the parent processes, the parent sandboxes must be kept alive to facilitate future warm starts.

**Restore MicroVMs from snapshots.** For MicroVM-based sandboxes, new instances are created by restoring from a previously captured memory snapshot rather than booting from scratch. MicroVMs that are idle do not need to be kept alive for long. Instead, they are snapshotted to serialized machine states and memory image. New invocation requests all enjoy the benefit of warm start as long as a snapshot for the required function image exists.

During the restoring process, the VM memory image is mapped directly and memory pages accessed during invocation are demand-paged in as the guest touches them. REAP [46] observes that functions access a stable working set of pages across invocations and builds a record-and-prefetch mechanism: it records the set of guest-physical pages touched during the first invocation and eagerly prefetches them from disk on subsequent restores, reducing cold-start latency. FaaSnap [12] further improves upon this by introducing a layered memory mapping method that turns major page faults requiring disk I/O on zero-initialized pages into minor page faults.

**Production systems use MicroVM-based sandboxes.** Modern production serverless platforms, such as AWS Lambda [14] backed by Firecracker [7], employ MicroVM-based sandboxes rather than cgroup-based containers for stronger isolation. Cgroup-based sandboxes share the host kernel: a single kernel vulnerability can compromise all co-located tenants on the same server. MicroVMs, by contrast, run a dedicated guest kernel inside a lightweight virtual machine monitor, presenting a much smaller and more stable host-guest interface. However, this stronger isolation boundary means that fork-based approaches do not directly apply to MicroVM sandboxes. Creating a new MicroVM cannot be reduced to remapping shared process states; rather, it requires allocating new kernel objects (including those managed by the KVM subsystem), instantiating separate virtual hardware resources (virtual CPUs, virtual network interfaces, virtual block devices), and configuring the guest address space, all of which are fundamentally per-VM constructs that cannot be shared across instances. As a result, snapshot-based restore remains the primary cold-start mitigation path for MicroVM-based systems, and improving the speed and efficiency of snapshot serving is the key challenge that Aquifer targets.

## 2.3 Motivation

### 2.3.1 Pool Capacity and Cost Tradeoff

The RDMA and CXL two pooling technologies occupy complementary points in the capacity–cost design space, and neither alone is sufficient for serving as a fleet scale solution.

RDMA-based pooling offers large capacity at low incremental cost. A single RDMA-capable NIC enables a node to reach stranded memory anywhere in the data center over the already-deployed Ethernet Clos network, requiring no exotic hardware beyond what is already in place. This makes RDMA the natural substrate for a cluster-wide memory pool with aggregate capacity in the tens of terabytes. However, as discussed in §2.1, every RDMA access incurs software overhead: page faults, kernel-level swapping, or application-level redirection, making it unsuitable for latency-sensitive page serving on the critical path of MicroVM restore.

CXL-based pooling eliminates this software overhead by exposing pool memory as a directly loadable/storable NUMA node, but its reach is fundamentally limited. Extending CXL connectivity beyond a pod would require CXL switching infrastructure: the first commercially available CXL switch to our knowledge, Structera S, was launched only in early 2026 by Marvell [3]. Deploying CXL switches at scale would mean adding a layer of CXL-specific switches alongside existing ToR and spine Ethernet switches, and interconnecting them with high-quality, long-range cables capable of meeting PCIe's stringent signal integrity requirements [35]. All of this is projected to be extremely costly, and cloud providers have explicitly cautioned against the assumption that CXL switching will be widely deployed in the near term [34].

Neither technology alone meets the requirements of fleet-scale snapshot serving: RDMA provides the capacity but not the latency, while CXL provides better performance but not the reach. However, the two tiers naturally compose into a hierarchical architecture. Figure 1 illustrates this design: within a pod, server nodes connect to a CXL v2.0 based MHD over short-reach PCIe links for low-latency, direct-access memory sharing; across pods, the existing Ethernet Clos network provides RDMA connectivity to a cluster-wide memory pool with larger capacity. Adopting this hierarchical design, however, introduces three new challenges that we address in the remainder of this section.

### 2.3.2 CXL Cache Coherence

The hierarchical design above places shared snapshot data in CXL MHD memory that is directly accessible by multiple hosts within the same pod. When multiple hosts concurrently restore MicroVM instances from the same snapshot, they would touch the same shared memory region. Ensuring the correctness of these concurrent accesses requires cache coherence: without it, a CPU may cache a stale copy of a cache line that another host has since modified, leading to silent data corruption or inconsistent state.

Hardware-managed cache coherence across hosts sharing a CXL device is specified only in CXL version 3.0 [1]. Realizing it requires coherence support across the entire memory access data path, including the host processors, any intermediate CXL switches, and the CXL memory device itself. However, the majority of existing processors and CXL memory hardware support only CXL v2.0 [13, 29], which provides no inter-host coherence guarantees [49]. Even looking forward, full hardware cache coherence across hosts has been reported to be economically unrealistic [29], as the silicon cost and complexity of maintaining a coherence directory at the device scales with the number of attached hosts and the size of the shared address space.

In the absence of hardware coherence, software must take responsibility for ensuring that concurrent accesses to shared CXL memory produce correct results. This requires carefully designing which data are placed in the shared region, how and when different hosts may read or write that data, and what mechanisms (e.g., cache-line flushes, memory fences, or access partitioning) are used to prevent stale reads and write conflicts. This presents our first challenge: ***how can we ensure the correctness of concurrent access to shared snapshot data in CXL memory without hardware cache coherence?***

#### 2.3.3 Snapshot Image Characterization

Existing MicroVM-based warm restore systems [12, 46] store the entire snapshot image, which contains a serialized full-size memory image of the VM instance. At fleet scale, this leads to a capacity challenge. For a sampled subset of a single Azure cloud region, there are 20,809 serverless applications as reported by [50]. Azure defaults to a minimum memory allocation of 1.5×GiB per function instance [4]. If at least one function in every application is snapshotted, this produces roughly 30 TiB of snapshot data in total. Storing all full-size images in disaggregated memory would be economically infeasible. This presents our second challenge: ***how should we redesign the snapshot format to minimize the storage overhead?***

To answer this, we conduct a characterization study of the composition of function snapshot images using 9 representative function workloads as shown Table 2 from FunctionBench [30] and SeBS [18]. We start each MicroVM from an initial snapshot taken after a cold boot, so that memory pages touched only during the one-time boot process are excluded from the characterization. We then iterate over different input data to issue 16 invocation requests per instance, reflecting the production scenario where 80% of invocation streak lengths are at most 16 (Figure 2). We use `userfaultfd` [41, 46] to serve page faults and record whether each accessed page was only read or also written to, using `userfaultfd` write protection.

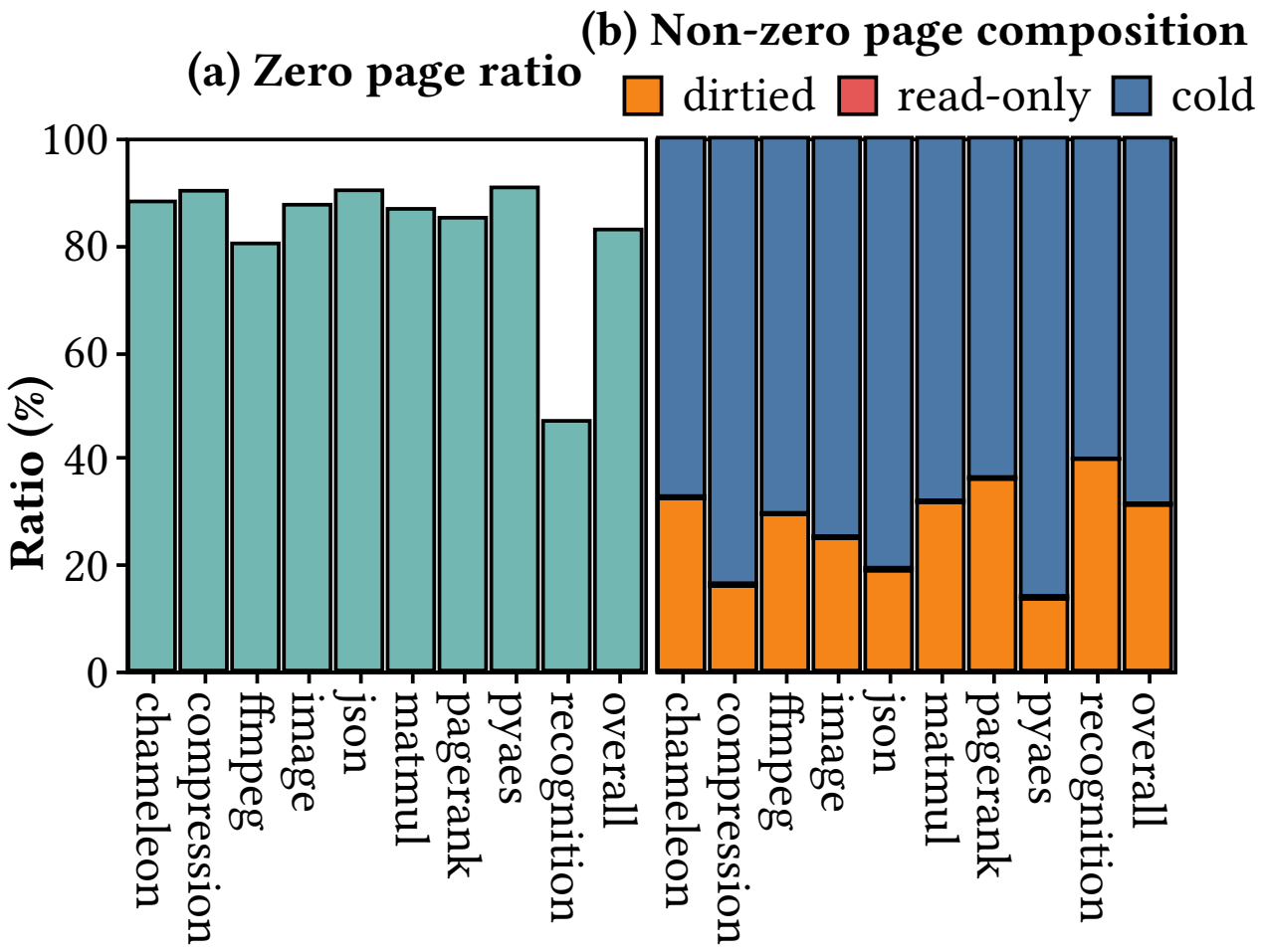

Figure 3: Composition of snapshot images across 9 representative serverless functions. Pages are classified into zero, cold, dirtied, and read-only categories. Each function instance is configured with 1.5×GiB of memory.

When taking the snapshot, we first walk all page contents to identify *zero pages*, pages whose contents are entirely zero. We then classify the remaining non-zero pages based on the recorded page fault information into three categories: (1) **cold pages**, non-zero pages that were never accessed during any of the 16 invocations; (2) **dirtied pages**, non-zero pages that were accessed and written to at least once, triggering a write-protection fault; and (3) **read-only pages**, non-zero pages that were accessed but never written to. We plot the breakdown in Figure 3.

The results reveal substantial redundancy in the snapshot image. Across all 9 workloads, on average 82.8% of memory pages are zero pages in the snapshot image, ranging from 46.9% of `recognition` to 90.7% of `pyaes`. These zero pages carry no useful state and need not be stored at all. Among the remaining non-zero pages, the majority are cold: on average, 72.7% of non-zero pages (60.2%–86.0% across workloads) are never accessed during the 16-invocation window. Only 5.4% of total pages are dirtied during invocations, and merely 0.05% are read-only. In other words, the pages that are actually needed for warm restore, the dirtied and read-only working set, constitute a small fraction of the snapshot.

These findings show that, by eliminating zero pages from the stored format, the 30 TiB aggregate shrinks to roughly 5.3 TiB. Furthermore, the cold pages, while non-zero, are not accessed during typical invocations and can tolerate the higher latency of the RDMA tier. Only the hot working set, dirtied and read-only pages, demands the low-latency access that the CXL tier provides. After partition applications based on the number of server pods available in the data center, each CXL pool only needs to handle ~130×GiB of data, assuming there are at least 10 server pods in a data center cluster.

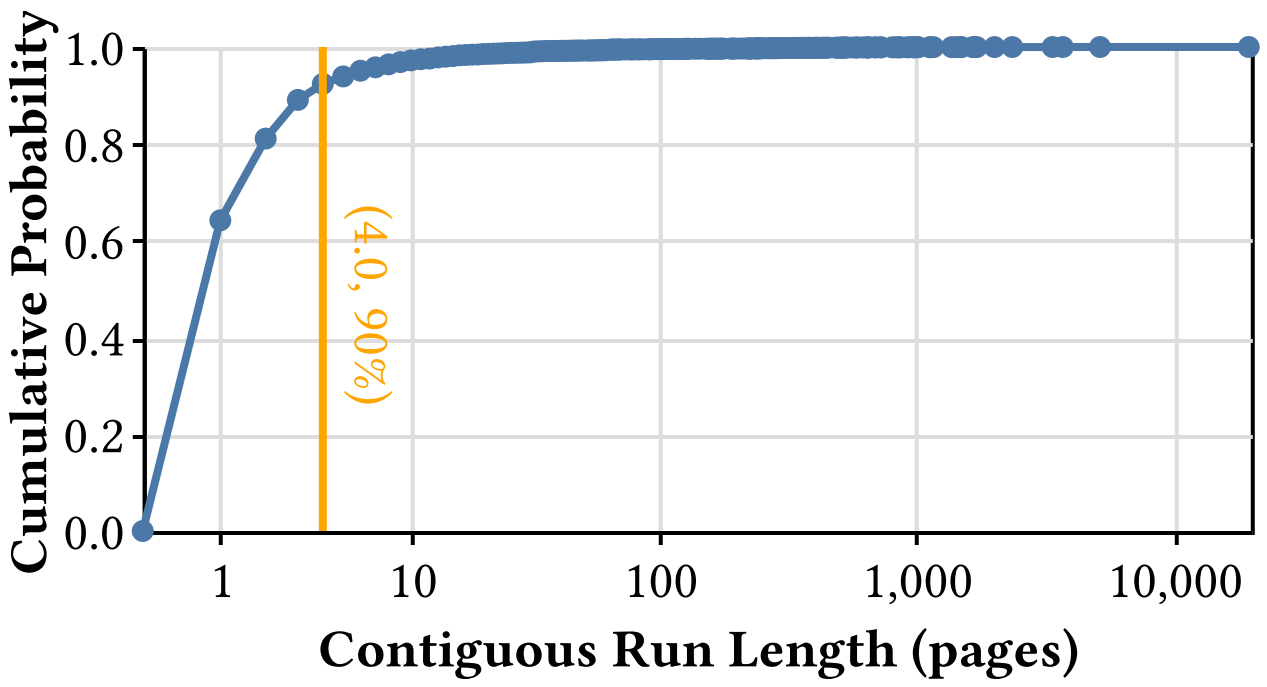

Figure 4: CDF of contiguous sub-range run lengths (in pages) within the hot working set across 9 representative serverless function snapshots. Over 90% of sub-ranges span fewer than 4 pages.

#### 2.3.4 Page Installation

The characterization above shows that any space-efficient snapshot format must eliminate zero pages and separate hot pages from cold ones. This necessarily means the stored image can no longer be a single contiguous image representing the full guest physical address space. This raises a third challenge: ***how can snapshot pages be efficiently installed into the MicroVM's address space when the snapshot image is compact and shared across concurrent restores?***

**Existing systems rely on contiguous, file-backed snapshots.** When a snapshot image is shared across concurrent MicroVM restores, it must remain immutable so that each instance observes identical initial state. FaaSnap [12] achieves this by storing a complete, contiguous snapshot as a regular file and `mmap`-ing the working set regions with read-write permissions while opening the file as read-only. The kernel's implicit copy-on-write (CoW) mechanism allocates a private page frame whenever the guest writes to a mapped region, leaving the original file pages untouched. However, this approach depends on two properties of the snapshot representation: the image is file-backed, so the kernel page cache provides CoW semantics; and it is contiguous, so a small number of `mmap` calls suffice to cover the address space. Any compact format that eliminates redundant pages will violate both properties.

**The working set is almost entirely written to.** Preserving immutability without file-backed CoW requires an alternative mechanism. A straightforward option would be to map snapshot regions as read-only and intercept writes via protection faults to create private copies. However, Figure 3 shows that dirtied pages account for 99.1% of the hot working set on average, while read-only pages contribute merely 0.05% of total pages. In other words, almost every page that a restored MicroVM touches will be written to. Mapping these pages as read-only would cause a protection fault on nearly every access, each of which must be caught by a user-space page fault handler, copied, and re-mapped with write permissions, an overhead proportional to the entire working set size.

**The working set addresses are highly fragmented.** Beyond immutability, efficiently mapping the working set requires that the hot pages occupy reasonably large contiguous regions, so that each `mmap` call installs many pages at once. To test whether the workloads exhibit this property, we measure the run-length distribution of contiguous address sub-ranges within the hot page set across all 9 function snapshots (Figure 4). Over 90% of contiguous sub-ranges span fewer than 4 pages, with an average run length of only 5.0 pages. Each snapshot contains an average of 4,164.2 such sub-ranges. This extreme fragmentation means that mapping the working set would require thousands of individual `mmap` system calls per restore. We measure the per-page cost of installing a page via `mmap` versus copying it with `uffd.copy()` even under the optimistic assumption that the `mmap`-installed page is directly accessible through CXL memory without involving any I/O, `mmap` is 2.6× slower per page. At thousands of sub-ranges per snapshot, the cumulative `mmap` overhead would dominate restore latency.

These observations motivate a page installation mechanism that avoids per-range `mmap` calls entirely, instead using `userfaultfd`-based copying to install pages into the guest address space on demand while keeping the shared snapshot image immutable.

## 3 Design

Section §2 identifies three challenges for serving MicroVM snapshots from a hierarchical memory pool: ensuring coherence without hardware support (§2.3.2), minimizing snapshot storage overhead (§2.3.3), and efficiently installing pages from a compact, non-file-backed image (§2.3.4). We first present an architectural overview, then address each challenge in the subsequent sections.

### 3.1 Architectural Overview

Figure 5 shows the major components of a serverless cluster running Aquifer. Aquifer introduces two components into the serverless infrastructure: ***orchestrators*** and a ***pool master***. Multiple orchestrators run on different server nodes that are connected to the same shared CXL and RDMA memory pools. A single pool master manages the pool-side storage of snapshot data. Aquifer reuses existing cluster-level serverless load balancers [6, 38] to distribute incoming function invocation requests across orchestrators. Each orchestrator acts as a MicroManager [14], managing the full lifecycle of MicroVM instances on its node: creating and restoring MicroVMs, serving page faults during execution, invoking the function endpoint, and returning results.

We now walk through the lifecycle of a function invocation to illustrate how the components interact.

**Request dispatch.** When an invocation request arrives at an orchestrator, and if there is not an already booted Mi-croVM instance for the requested function, the orchestrator must create a new instance.

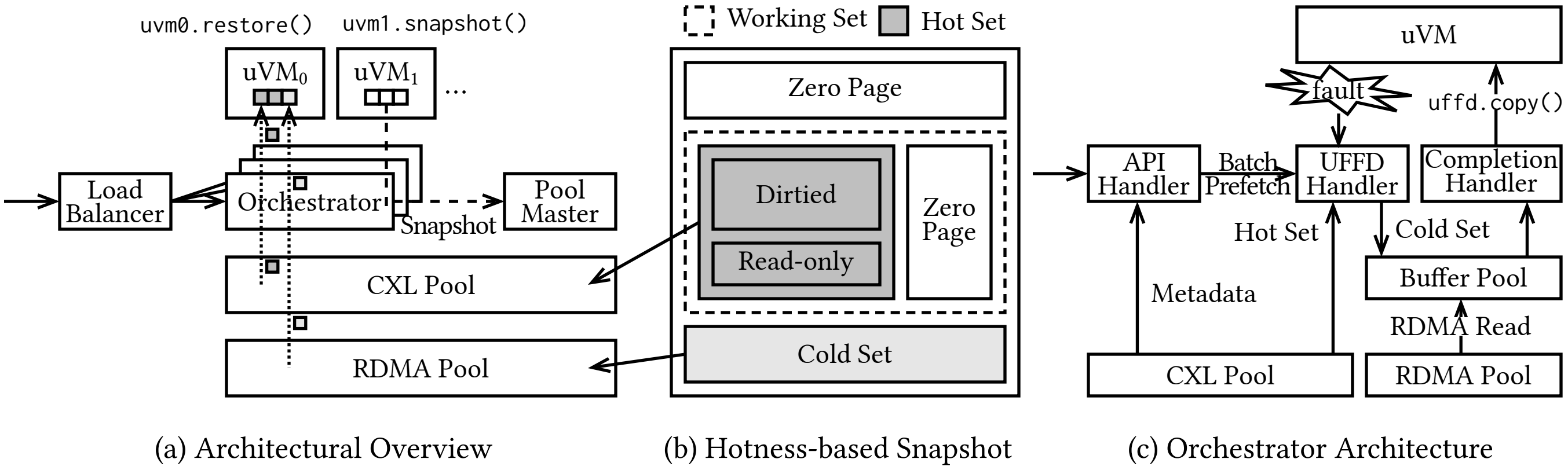

(a) Architectural Overview (b) Hotness-based Snapshot (c) Orchestrator Architecture

Figure 5: Design

**Cold boot or warm restore.** If a snapshot is available for the requested function image exists in the memory pool, the orchestrator performs a warm restore: it claims a skeleton MicroVM from pre-created ones (§3.5) and maps the guest memory region with `userfaultfd` registered, so that page faults during execution can be served from the pool. The orchestrator then loads the serialized machine state, such as vCPU registers, and resumes the MicroVM. If no snapshot exists, the orchestrator falls back to a full cold boot, after which it takes an initial snapshot for future restores.

**Page fault serving.** Once the restored MicroVM begins executing, guest memory accesses to pages not yet present in local memory trigger `userfaultfd` page faults. The page serving mechanism is described in §3.4.

**Snapshot creation.** MicroVM instances that are explicitly requested by the platform are snapshotted. When creating a snapshot, the orchestrator forwards the MicroVM's memory and machine state to the pool master for storage (§3.2).

Throughout the lifecycle, orchestrators loading snapshot pages from the shared pool must observe a consistent view of the data, even as the pool master concurrently accepts new or delete snapshots from other nodes. The coherence protocol that ensures correctness is described in §3.3.

### 3.2 Hotness-based Snapshot Format

To address the third challenage in §2.3.3, Aquifer proposes a hotness-based snapshot format which eliminates redundant data and partitions the remainder by hotness so that each subset can be placed in the appropriate pool tier.

**Eliminating zero pages.** Our characterization shows that on average 82.8% of pages in a snapshot image are all zero. These pages carry no useful state: when a restored MicroVM faults on a guest physical address whose snapshot content is zero, the orchestrator can simply install a freshly zeroed page without consulting the pool. Aquifer therefore excludes zero pages from the stored snapshot entirely.

**Hot–cold partitioning.** Among the remaining non-zero pages, the characterization reveals that the vast majority are cold, never accessed during the profiling window. Only the hot working set, dirtied and read-only pages, is accessed during typical invocations. Aquifer partitions non-zero pages into two subsets based on this classification: hot pages are stored in the CXL pool, where they can be served via direct load/store, and cold pages are stored in the RDMA pool, which provides the capacity needed at lower cost.

**Offline Hotness profiling.** Aquifer generates snapshots offline to avoid imposing profiling overhead on production invocations. To authentically reflect the access patterns of real applications, the orchestrator samples invocation requests previously made to the targeted function image and replays them in a freshly restored MicroVM. For each page served by the orchestrator during replay, the faulting address is recorded into a *working set* array. Since our characterization in §2.3.3 shows that read-only pages are negligible (0.05% of total pages), Aquifer does not need to distinguish between read and write accesses as the characterization study does; it only needs to track which pages are accessed at all. After iterating over the sampled invocations, the orchestrator takes the snapshot and derives the page sets. It first walks the memory state to identify zero pages. The hot set is then computed by excluding zero pages from the recorded working set, and the cold set consists of all non-zero pages not in the working set.

**Snapshot layout.** The pool master is responsible for storing the final snapshot. Each snapshot produces a **catalog entry** and two **data regions**, one in CXL memory and one in RDMA memory. The catalog entry records the function image name and the offsets to the two data regions. Each data region contains compacted page content: only the non-zero pages classified as hot or cold, respectively, are stored, so the data within each region are not contiguous in the guest address space.

To translate a guest page address to the corresponding offset within its data region, Aquifer maintains a page-table-like *offset array* with one slot per guest page address. Each slot stores the offset into the appropriate data region. To distinguish whether a page resides in CXL or RDMA memory, the top unused bits of each slot encode the memory backend type, avoiding a separate data structure for tier lookup. Zero pages have no entry and are identified by a sentinel value. The offset array, together with the serialized machine state, is stored alongside the hot page data in CXL memory, so that orchestrators can access both the index and the machine state directly over load/store during restore without incurring RDMA round trips on the critical path.

### 3.3 Coherence Protocol

To address the second challenge from §2.3.2, we build an ownership-based coherence protocol using atomic operations, which we assume are supported despite the lack of inter-host cache coherence [49].

**Ownership model.** All snapshots in the CXL pool are owned by the pool master, which is the sole entity responsible for publishing new snapshots, updating existing ones, and deleting stale entries. Once a snapshot is published, it may be concurrently accessed by multiple orchestrators, which act as *borrowers*. Borrowers are only permitted to read snapshot data; they never modify it. This immutability invariant is the foundation of the protocol: because snapshot data is never written by borrowers, the coherence problem reduces to ensuring that borrowers always read up-to-date data and that the owner does not mutate data while any borrow is active.

**Catalog entry metadata.** Each snapshot catalog entry in CXL memory is augmented with two fields: a *state* field and a *refcount*. The state is either `published` (the snapshot is published and available for borrowing) or `tombstone` (the snapshot is marked for deletion or update and no new borrows are permitted). The refcount tracks the number of orchestrators currently borrowing the snapshot. For a newly published snapshot, the state is set to `published` and the refcount to zero.

**Borrow protocol.** When an orchestrator needs to restore a MicroVM from a snapshot, it first atomically increments the refcount of the corresponding catalog entry, then issues a compare-and-swap on the state field expecting `published`. If the CAS succeeds, the state remains `published` and the borrow is established. If the CAS fails (the state is `tombstone`), the snapshot is being reclaimed; the orchestrator atomically decrements the refcount and falls back to a cold start. Incrementing the refcount before the state CAS ensures there is no window in which the pool master could observe a zero refcount while an orchestrator is in the process of borrowing. The CAS on the state field provides an atomic verification that is properly ordered with respect to the preceding refcount increment.

After a successful borrow, the orchestrator must ensure it does not read stale data that may be cached from a previous version of the same catalog entry or data region. It invalidates the relevant cache lines using `clflushopt`, covering the machine state, offset array, and page content. Subsequent loads are guaranteed to fetch the current data from CXL memory. Once the orchestrator finishes serving the MicroVM and shuts it down, it atomically decrements the refcount.

**Owner protocol.** When the pool master needs to modify or remove a snapshot, it atomically sets the state to `tombstone`, which prevents any further borrows from succeeding. The subsequent action depends on the operation:

*Deletion:* the pool master sets the state to `tombstone` and the entry is considered logically deleted. The underlying data region is reclaimed only after the refcount drops to zero, ensuring that any in-flight borrows complete without accessing freed memory.

*Update:* the pool master sets the state to `tombstone` and waits for the refcount to drop to zero, ensuring all active borrowers have finished. It then writes the new snapshot data to the CXL and RDMA data regions and sets the state back to `published` with the refcount at zero, publishing the updated snapshot.

*Add (reuse):* when publishing a new snapshot, the pool master selects a catalog entry in `tombstone` state whose refcount has reached zero, writes the new snapshot data, and sets the state to `published`.

### 3.4 Page Serving

As shown in §2.3.4, the compact snapshot format and the CXL pool's non-file-backed address space rule out conventional `mmap`-based page installation. Aquifer instead installs all pages via `userfaultfd`'s `uffd.copy()` interface. When a guest memory access targets a page not yet present, the `userfaultfd` mechanism delivers the fault to the orchestrator, which looks up the faulting page frame number in the offset array. The top bits of the corresponding entry indicate which memory backend, either CXL or RDMA, holds the page, and the page address gives the location within that backend's data region. Because `uffd.copy()` creates a private copy of the page in the guest's address space, the pool-resident snapshot data is never modified, preserving immutability across concurrent restores without requiring file-backed CoW or read-only mappings.

**Hot set pre-installation.** Before resuming the restored MicroVM, the orchestrator pre-installs the entire hot set recorded during offline profiling. For each hot page, the orchestrator reads the page address from the offset array and issues a `uffd.copy()` with the CXL pool memory as the source. Because CXL memory is directly accessible via load/store, the kernel can copy directly from the CXL data region into the guest address space without an intermediate buffer. Pre-installing the hot set eliminates page faults for the vast majority of accesses during execution, allowing the restored MicroVM to run at near-native speed from the start.

**Demand-paging cold pages.** Cold pages are not pre-installed. Our characterization shows that these pages were not accessed during any of the 16 profiling invocations, so the probability of access during a production invocation is low. Eagerly prefetching cold pages would add restore latency for negligible benefit. Instead, cold pages are served on demand only when the guest actually faults on them.

**Asynchronous RDMA fault handling.** When a fault does arrive for a cold page, the orchestrator must fetch it from the RDMA pool. To minimize the latency impact, Aquifer uses an asynchronous architecture that decouples fault handling from RDMA completion. The orchestrator maintains a local buffer pool of free pages. Upon receiving a cold-page fault, the fault handler acquires a buffer page from the pool and issues an asynchronous RDMA read into that buffer, then immediately returns to process the next pending fault. A separate completion handler thread polls the RDMA completion queue. When a read completes, the completion handler installs the fetched page into the guest address space via `uffd.copy()`. This split avoids head-of-line blocking: the fault handler is never stalled waiting for an RDMA transfer to finish, and multiple RDMA reads can be in flight concurrently, overlapping network latency with continued fault processing.

### 3.5 MicroVM Pool

Every MicroVM requires a set of host resources before it can start: a jailer cgroup and namespace for isolation [14], a Firecracker VMM process [7], virtual network interfaces, bridges, and other virtual devices. Allocating these resources on the critical path adds significant control-plane latency [15]. To avoid this, each orchestrator maintains a MicroVM pool of pre-created skeleton instances with all host resources already provisioned. The pool is continuously replenished in the background, so that when a new MicroVM is needed, a skeleton instance can be claimed immediately.

### 3.6 Limitations and Discussion

*CXL atomics.* Aquifer's coherence protocol (§3.3) relies on atomic operations across hosts sharing a CXL MHD, an assumption supported by prior empirical findings [49] but not formally guaranteed by the CXL 2.0 specification [1]. For CXL pools that do not support cross-host atomics, correctness can be preserved by falling back to an RDMA-based RPC: the orchestrator queries the pool master to acquire a lease on the snapshot before restore and issues a second RPC upon MicroVM shutdown to release it. Because all orchestrators communicate with the pool master, it can guarantee that no concurrent update or deletion occurs while active leases exist, deferring such operations until all leases expire.

*CXL pool eviction.* The current design assumes the CXL pool always has sufficient capacity to store all hot snapshot data. In production, this may not hold as the number of active function images grows. To support efficient CXL pool utilization, we can augment each snapshot catalog entry with a *borrow counter* alongside the existing refcount. The pool master periodically collects and resets the borrow counters to build an eviction candidate list ranked by both temporal locality and access frequency. When CXL memory is under pressure, the pool master deletes the lowest-ranked snapshots from recent collection cycles, freeing CXL capacity for newly created snapshots.

*Hotness shifting.* Aquifer profiles the hot working set offline from a fixed sample of invocations (§3.2). If a function's access pattern shifts over time, for example due to changes in input distributions, the recorded hot set may become stale, causing CXL-tier misses and additional RDMA-tier faults on the critical path. This can be mitigated by introducing a periodic refresh: after a configurable interval, Aquifer resamples recent invocations and regenerates the snapshot with an updated hot/cold partition.

*Fault tolerance.* Aquifer employs a single pool master per CXL memory pool. Although this constitutes a single point of failure, its impact on the critical invocation path is minimal. Orchestrators restoring from an already-published snapshot do not need explicit communication with the pool master; they follow the coherence protocol to update the snapshot catalog and proceed with restore independently. Only snapshot modifications require pool master involvement, and these operations are infrequent relative to invocation traffic. Furthermore, because all snapshot data resides in the shared CXL and RDMA pools and is visible to every connected node, a replacement node can be manually elected as the new pool master and resume normal operation. To automate failover, a consensus protocol such as Raft [40] can be introduced among the nodes in the same pod: when heartbeats from the pool master cease, a new leader is elected and takes over pool management.

*Snapshot deduplication.* Many serverless functions share common runtime components (e.g., Python interpreter, shared libraries), leading to duplicate pages across per-function snapshots. Because snapshot pages are indexed via the offset array, which can point to any valid CXL or RDMA pool address for a given guest page, a deduplication layer can be transparently integrated into the pool master and applied when creating new snapshots. However, deduplication affects the restore path. In the current design, the orchestrator issues `clflushopt` over a contiguous CXL data region, which is relatively inexpensive. With deduplicated pages scattered across the pool, the orchestrator must instead walk the offset array and flush each individual page address. Moreover, flushing a shared page may evict a cache line currently in use by another MicroVM on the same host, causing performance interference.

*Security and isolation.* Aquifer's shared CXL memory pool is accessible to all connected orchestrator and pool master nodes. We assume these host-side components are trusted and that untrusted code is confined within MicroVM sandboxes, consistent with the threat model of production serverless platforms [7, 14]. MicroVMs provide a strong isolation boundary: each instance runs behind a dedicated guest kernel and a lightweight VMM, presenting a minimal host–guest interface. On the host side, Firecracker's jailer provides a secondary containment facility, placing each MicroVM process in a separate namespace and cgroup context to limit the blast radius of a potential VMM compromise.

# 4 Implementation

We implement the Aquifer orchestrator and pool master from scratch in Rust, totaling 5,953 lines of code. We use the most recent vanilla Firecracker v1.15 as the MicroVM monitor without modification. Both host and guest run Ubuntu 24.04 with its stock Linux 6.8.0 kernel. On the guest side, we set the kernel command line parameter `init_on_free=1`, which zeroes pages in the kernel page allocator upon freeing. This ensures that snapshot images do not contain stale data from lazily sanitized pages [12].

**Page fault handling.** Firecracker v1.15 natively supports `userfaultfd`-based snapshot restore: upon receiving a restore request, Firecracker registers the guest memory region with `userfaultfd` and passes the resulting file descriptor to the orchestrator daemon, which becomes the external page fault backend. The orchestrator multiplexes fault events from all active MicroVMs on a single dedicated thread using an `epoll`-based event loop. On the fast path (§3.4), zero-page faults are resolved via `uffd.zeropage()` rather than `uffd.copy()`, avoiding an unnecessary memory copy. The RDMA completion handler (§3.4) uses a hybrid polling strategy: upon receiving a completion it busy-polls for up to 1,024 iterations to batch further completions; if none arrive, it yields the CPU and switches to the RDMA completion channel for event-driven notification on the next completion event delivered by the NIC.

**CXL memory interface.** Aquifer exposes CXL memory to each node through the kernel DAX subsystem as a device file under `/dev/dax`. The DAX subsystem is involved only in the control-path lifecycle management of the device; once the device is configured, subsequent load/store accesses bypass the kernel entirely. To access CXL pool memory, the orchestrator or pool master simply calls `mmap()` on the DAX device file, after which the CXL memory region is accessible as ordinary virtual memory.

# 5 Evaluation

## 5.1 Evaluation Methodology

### 5.1.1 Environment Setup

We conduct all experiments on CloudLab [20]. Because CloudLab does not currently offer machines with CXL-attached memory devices, we emulate CXL memory using a remote NUMA node, following prior work [35]. We use a dual-socket r6525 node from the CloudLab Clemson cluster, shown in Table 1.

| Component | Specification |
|---:|---|
| CPU | 2× AMD EPYC Milan 7543 @2.8 GHz, 32 cores |
| DRAM | 2× 8× 16 GiB ECC DDR4-3200 |
| System Disk | 480 GiB Samsung PM883 SATA SSD |
| Data Disk | 1.6 TiB Samsung PM1733 NVMe SSD |
| Control NIC | 25 Gb/s Mellanox ConnectX-5 MT27800 |
| Experiment NIC | 100 Gb/s Mellanox ConnectX-6 Dx MT28924 |

Table 1: Hardware configuration of CloudLab r6525 nodes.

**CXL pool emulation.** To emulate a shared CXL memory pool on a single dual-socket machine, we run two VMs on one NUMA node, one acting as the orchestrator and the other as the pool master, each pinned to 16 dedicated physical cores with 64 GiB of memory. The memory attached to the remote NUMA node serves as the CXL pool media, while emulating the access latency characteristic. Each VM attaches the emulated CXL region through a `virtio-pmem` device [5], which exposes the region as a DAX device. We use `virtio-pmem` solely for device enumeration; all data accesses and cache flushes operate directly on the `mmap`'d memory region without going through the `virtio-pmem` flush command. The entire emulated CXL region is mapped into both VMs, ensuring that both nodes observe the same pool contents. Although the two VMs share the same physical machine and thus have implicit cache coherence over the backing memory, we still treat the region as non-cache-coherent and require `clflushopt` on every borrow, emulating a CXL 2.0 MHD where no inter-host coherence is provided.

**RDMA networking.** Because only one NIC is available for experiments on each physical node, we use SR-IOV [19] to create one 100 Gb/s virtual function per VM. The virtual functions are interconnected through the embedded switch (eSwitch) of the ConnectX-6 Dx NIC, providing each VM with an independent RDMA-capable interface. eSwitch feature can only be enabled in modern driver and firmware versions, thus we use the DOCA driver version 2.9.3 and firmware version 22.47.1088. The pool master's local DRAM serves as the backing memory for the RDMA pool.

### 5.1.2 Workloads

We evaluate Aquifer using 9 serverless function workloads drawn from FunctionBench [30] and SeBS [18], spanning web serving, multimedia processing, scientific computing, and machine learning. Table 2 summarizes each workload and its application domain.

### 5.1.3 Comparison Targets

To the best of our knowledge, Aquifer is the first system to exploit a two-tier CXL+RDMA memory pool for MicroVM snapshot serving, so no direct comparison target exists. Instead, we adapt prior MicroVM snapshot restore techniques to operate over the same RDMA memory pool used by Aquifer, ensuring that performance differences reflect

| Workload | Domain | Description |
|---|---|---|
| chameleon | Web | HTML table rendering |
| compression | Web | File compression |
| json | Web | JSON serialization/deserialization |
| ffmpeg | Multimedia | Video greyscale filter |
| image | Multimedia | Image rotation |
| matmul | Scientific | Matrix multiplication (NumPy) |
| pagerank | Scientific | Graph PageRank (igraph) |
| pyaes | Scientific | AES encryption |
| recognition | ML | Image recognition (ResNet) |

Table 2: Serverless function workloads used in evaluation.

algorithmic design choices rather than storage media. We compare the following five configurations:

*Firecracker:* The baseline demand-paging configuration. No prefetching is performed: every guest memory access to a missing page triggers a `userfaultfd` fault, which the handler serves by issuing an RDMA read to fetch the page from the pool and installing it via `uffd.copy()`.

*REAP:* An adaptation of the record-and-prefetch approach [46] from local file storage to the RDMA memory pool. In our adaptation, the prefetch phase issues RDMA reads for all working-set pages and installs them via `uffd.copy()`. Pages outside the working set are demand-paged as in *Firecracker*.

*FaaSnap:* FaaSnap [12] only prefetch non-zero-page in the working set by an overlay-based memory mapping scheme. Our adaptation, the prefetch phase issues RDMA reads for the hot set pages, which contains no zero-page, and installs them via `uffd.copy()`, reducing the number of RDMA reads compared to REAP.

*FcTiered:* Extends the *Firecracker* baseline with Aquifer's hotness-based snapshot format (§3.2) and two-tier page serving. Hot pages are served directly from the CXL pool via `uffd.copy()` without RDMA, while cold pages fall back to RDMA-based demand-paging. No prefetching is performed.

## 5.2 Ablation Study

To understand Aquifer's performance, we perform an ablation study on the chameleon workload, as shown in Figure 6 (a) and (b), measuring time spent in each stage of the invocation process across all five configurations. Before the guest OS resumes in the MicroVM to execute the function, the orchestrator performs a setup sequence: it prepares the machine state, configures the `userfaultfd`-based memory backend by calling the Snapshot API, handshakes with Firecracker to obtain the page fault handling context, prefetches pages, and finally resumes the MicroVM. During function execution, we measure the time spent in `uffd.copy()` as a proxy for the major part of the page fault installation cost, as shown in Figure 6 (c).

Aquifer achieves 2.12× lower total invocation time than Firecracker and 1.19× lower than FaaSnap. In the setup stage, Aquifer is 1.45× and 1.67× faster than FaaSnap and REAP, respectively. Firecracker and FcTiered skip the prefetch phase entirely, resulting in minimal setup time, but they pay a steep price during execution: their page installation cost is 187× higher than Aquifer's, as every guest memory access triggers a demand-paged fault. FaaSnap and REAP incur higher Snapshot API and page installation costs than Aquifer because they lack the hotness-based snapshot format and must issue more RDMA reads, which are significantly slower than serving a page directly from CXL memory, when Firecracker accesses the guest memory range. Aquifer's prefetch phase is slower than FaaSnap's and REAP's because CXL-based page pre-installation is currently serialized, whereas RDMA requests can have many operations in-flight concurrently, allowing the RNIC to leverage multiple Direct Memory Access (DMA) engines for parallel remote memory reads.

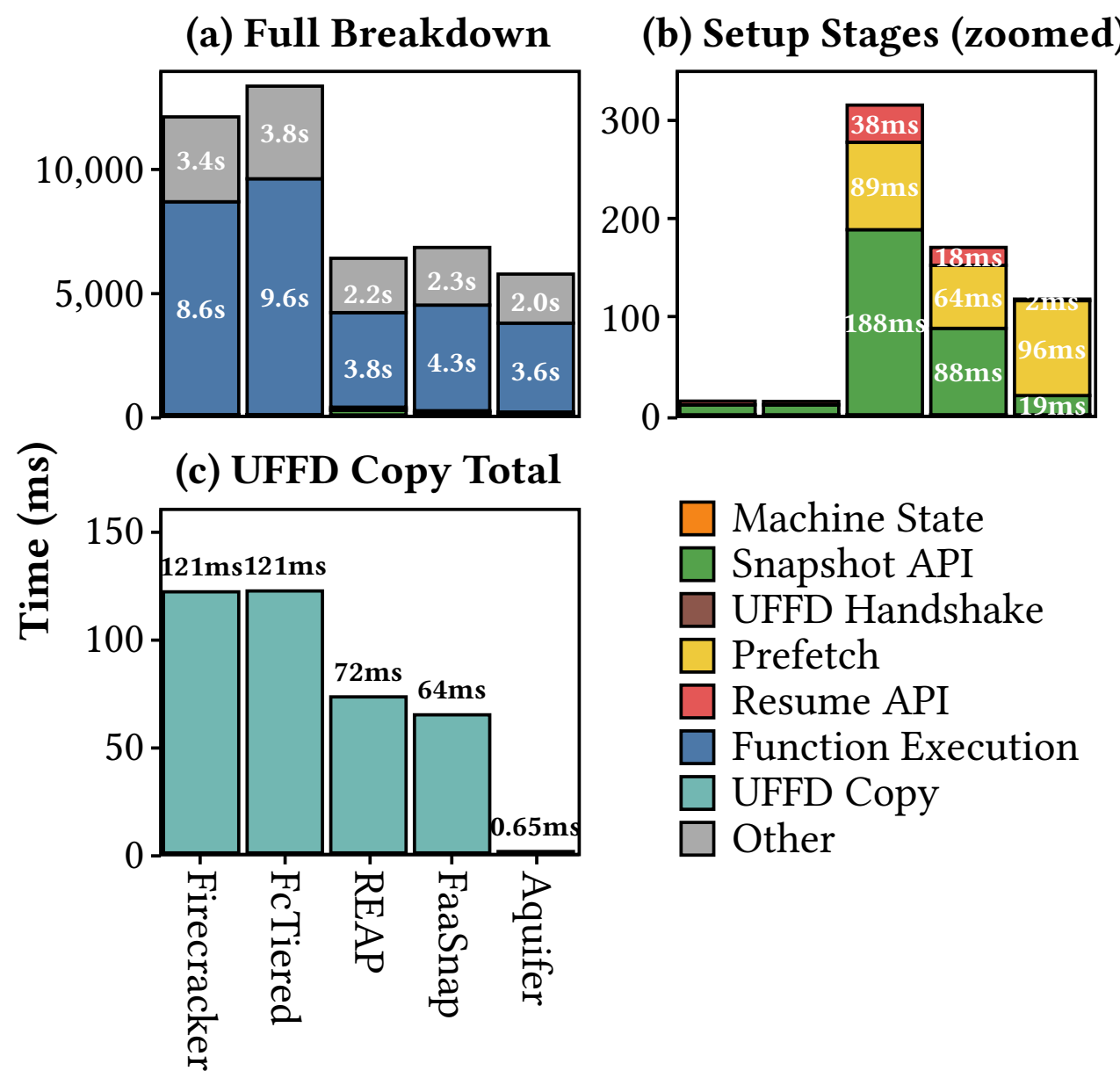

Figure 6: Invocation time breakdown for chameleon, averaged across 32 concurrently restored instances.

## 5.3 Scalability

To study how Aquifer scales with increasing concurrency, we restore $n$ MicroVMs simultaneously, sweeping $n$ in a semi-logarithmic progression (1, 2, 4, 8, 12, 16, 24, 32) across all nine function workloads. The upper bound of 32 is a limitation of our current benchmarking harness, not of the system itself. The recognition workload only scales to 16: its compute-intensive, long-running nature causes invocations at higher concurrency to exceed the connection timeout, after which the orchestrator kills them.

Figure 7 shows that Aquifer is the fastest configuration in eight of nine workloads. The sole outlier is ffmpeg, where REAP outperforms Aquifer. This is because ffmpeg extensively writes intermediate files to tmpfs and later frees them, creating a large number of zero pages in the working set.

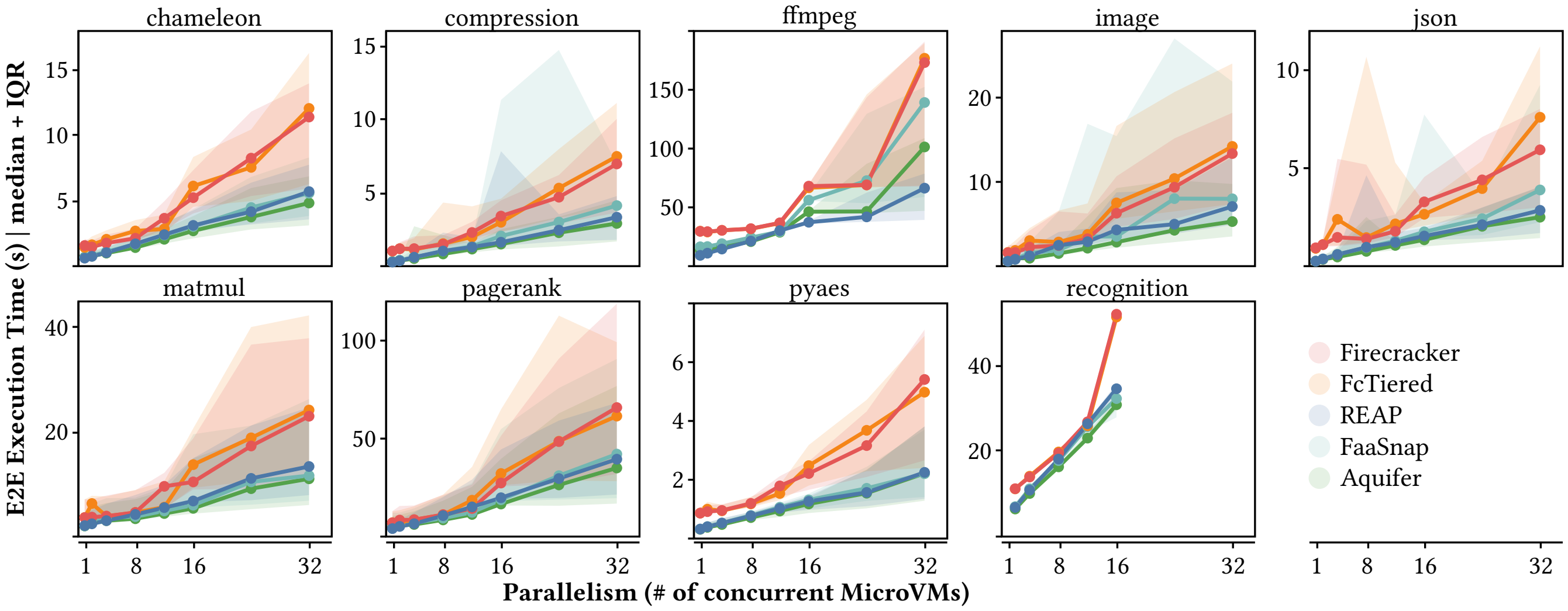

Figure 7: End-to-end invocation execution time (median and interquartile range) as the number of concurrently restored MicroVMs increases from 1 to 32, across all 9 function workloads.

REAP prefetches the entire working set including these zero pages, whereas Aquifer only prefetches non-zero pages, incurring more demand-paged faults during execution and increasing overall invocation time. Overall, Aquifer achieves a geometric-mean speedup of 2.2× over Firecracker and 1.3 × over FaaSnap, with per-workload speedups over FaaSnap ranging from 1.00× under pyaes to 1.57× under json. For pyaes, as this workload is compute-centric, the amount of memory touched is relatively small, so the resulting performance of FaaSnap and Aquifer is close. Compared to REAP, Aquifer's geometric-mean speedup across all nine workloads is 1.1×; excluding ffmpeg, per-workload speedups range from 1.02× under pyaes to 1.35× under image.

## 6 Related Works

*CXL memory in virtualized clouds:* Memstrata [51] demonstrates a hardware-based CXL tiering system using Intel Flat Memory Mode and designs a memory allocator to reduce inter-VM contention. Demeter [26] argues for delegating tiered memory management to the guest kernel, which is closer to locality information and has vPMU support, to reduce TLB overhead and improve scalability.
*Hybrid far memory systems:* Atlas [16] simultaneously supports page-based and object-based far memory access, dynamically switching between the two based on a tradeoff of data locality and I/O amplification. Ditto [44] enables real-time switching among different caching algorithms to improve the hit rate of a far memory caching system.
*Coherence and fault tolerance:* Tan et al. [45] formally model and verify the CXL.cache protocol, proposing clarifications and fixes to the specification. DRust [37] shows that the ownership model in Rust can be leveraged to simplify coherence in RDMA-based distributed shared memory systems. SWARM [39] designs a replication scheme for shared objects in disaggregated memory. Aceso [27] improves fault tolerance of key-value stores using a hybrid checkpointing and erasure coding scheme tailored for index and KV data.
*Serverless snapshot compression:* Sabre [33] uses the Intel In-Memory Analytics Accelerator to speed up compression and decompression of MicroVM snapshot images. Snapipeline [32] pipelines snapshot decompression with memory page restoration and function execution.

## 7 Conclusion

We presented Aquifer, a hierarchical memory pooling system that combines pod-local CXL memory with cluster-wide RDMA memory to serve MicroVM snapshots in a hierarchical pool. Our characterization of snapshot images revealed that the vast majority of pages are either zero or cold, motivating a compact, hotness-based snapshot format that partitions pages across the two tiers by access information. To ensure correctness on CXL 2.0 hardware lacking inter-host cache coherence, Aquifer employs an ownership-based coherence protocol built on atomic operations. A copy-based page serving mechanism pre-installs hot pages from CXL memory and demand-pages cold pages asynchronously from RDMA. Aquifer achieves a 2.2× geometric-mean invocation speedup over Firecracker and 1.3 × over FaaSnap across nine serverless workloads, demonstrating that hierarchical CXL+RDMA pooling is a practical substrate for latency-sensitive cloud workloads.